
\magnification=1300
\baselineskip=14pt

\rightline {OITS-541}
\rightline {April 1994}
\vskip1.5cm
\centerline{\bf  CLUSTER GROWTH IN TWO-DIMENSIONAL}
\medskip
\centerline{\bf QUARK-HADRON PHASE TRANSITION}
 \vskip1.5cm
\centerline{\bf Rudolph C. Hwa$^1$\ \  {\rm and}\ \  Jicai\ Pan$^2$}
\bigskip
 \centerline{{$^1$}Institute of Theoretical Science and Department of
Physics}

\centerline{University of Oregon, Eugene, Oregon 97403}
\medskip
\centerline{{$^2$} Department of Physics, McGill University}
\centerline{ Montreal, Quebec, Canada  H3A 2T8}

\vskip1.5cm
\centerline{\bf Abstract}
\vskip.8cm
The problem of hadronic cluster production in heavy-ion collisions is
studied in search for an observable signature of first-order quark-hadron
phase transition. The study is carried out by cellular automata  in a
two-dimensional model of the mixed phase at midrapidity. The clusters are
allowed to grow as well as to coalesce upon collision. The distribution of
cluster sizes is found to exhibit scaling behavior that is independent of
the size of the mixed region, nucleation radius and nucleation probability.
The universal scaling index $\gamma=1.86\pm 0.18$ may be used to
characterize and identify the phase transition process. Possible connection
with self-organized criticality is pointed out.

\vfill
\eject

\noindent {\bf 1. INTRODUCTION}
\medskip
In a recent paper we have initiated the study of cluster production
in quark-hadron phase transition in heavy-ion collisions [1]. The purpose
is to investigate the possibility of the existence of new and interesting
features that characterize the first-order phase transition that are
independent of the details of the hadronization process, which may be so
complicated as to render any feasible analytical approach too
oversimplified.   We used a cellular automaton to show the existence of
a scaling behavior in the sizes of the hadronic clusters produced.
Although the study is susceptible to the criticism in that the rules
adopted are also an oversimplication of the phase transition process, the
focus has been shifted from finding an analytical description of the
problem on the basis of ignoring certain complications to focusing on those
complications in the framework of some simple rules.  Specifically, our
emphasis has been on the nature of fluctuations of the cluster sizes by
taking into account cluster collisions, coalescence and breakup during the
growth process.  In [1] we consider only the 1-dimensional problem, which
is a drastic simplification of the realistic case.  In this paper we make
the first generalization to two dimensions, and examine the associated
complexity.  We shall show the persistence of the scaling behavior and
investigate the origin of the phenomenon.
\medskip
We mentioned in [1] the possible connection between the cluster growth
problem and self-organized criticality (SOC) [2]. If SOC is defined as the
behavior of an extended system that exhibits nontrivial scaling properties
without the tuning of a parameter, then we have found that under certain
general conditions the hadronization of a quark-gluon plasma possesses the
characteristics of SOC.  However, it is not clear what the nature of the
critical point is. It is not an alternative, nonthermal description of a
system going through what is usually regarded as a first-order phase
transition of a thermal system. We have not understood the phenomenon well
enough to connect the critical behavior to some features in continuum
mechanics, such as a singularity in the diffusion constant for the sandpile
problem [3].  What we have found are some interesting
properties of heavy-ion collisions that are worthy of more extensive
examination from various points of view.
\medskip
In two dimensions (2D) there are new features that are absent in 1D.
The most prominent one is that the clusters can have various shapes.
Depending on the detail dynamics, such as the positivity and magnitude of
surface tension, different types of shapes may develop as the clusters
grow.
The probabilities of coalescence and breakup will be affected; they in turn
can influence the cluster shapes. Clearly, the more dynamical features are
to be included in the modeling, the more complicated the rules of the
cellular automaton will have to be. The problem will then become one
opposite in approach to what are usually done in simulations for SOC, the
aim of which is to adopt the simplest set of rules that can generate a
phenomenon that exhibits the SOC behavior.  For example, for the sandpile
problem [2] or the forest fire problem [4, 5] the models are not close to
reality but are nevertheless interesting because each of them exemplifies a
class of problems that breaks new ground in statistical physics. Our
problem
here is somewhere in between.  We are interested in a specific problem:
that
is, the discovery of observable signatures of quark-hadron phase transition
in heavy-ion collisions. The relevant dynamics that controls the nature of
fluctuations at the transition point is not known well enough to specify
the rules of cluster growth. Indeed, the aim is to search for possibly
interesting features that are independent of the details of the rules, and
if those features exist in the modeling, then there will be strong
motivation
to look for them in the experiments. What are interesting about
fluctuations
are scaling and universal properties, which should emerge from simple
models, not complicated ones, if the results are to be taken seriously as
being generic.  Thus we must, at least in the beginning, try to be as
simple
as possible, while maintaining as much as possible the essential
ingredients
of the growth process that characterize the heavy-ion collisions.
\medskip
In the following we shall for the sake of simplicity ignore the
possibility of breakup of the clusters on their way out of the plasma. In
the 2D case it is possible to devise many ways to break up a cluster,
with or without collisions. After examining a few possibilities, we have
decided to leave out breakup altogether, in order not to confuse the issue
of what is responsible for the results obtained without breakup.
\medskip
In a realistic collision problem there are many parameters that specify
the system undergoing phase transition, e.g., size of the mixed phase,
transverse expansion velocity, critical radius for nucleation, nucleation
rate, growth rate, etc.  Varying all those parameters will complicate the
problem so much that the essential characteristics of the scaling behavior
will easily get lost. We shall therefore vary a few parameters only to see
the sensitivity of the result. We emphasize that those parameters are not
varied in order to bring the system to the critical point, but for the
purpose of exploring the universality of the result. That is a very
important
difference.
\vskip.7cm
\noindent {\bf 2. CLUSTER FORMATION}
\medskip
To pose the physical problem we adopt the conventional global picture of
the collision of two high-$A$ nuclei at very high energy. After impact
there is rapid longitudinal expansion with slower radial expansion.
Focusing on a thin slice of the expanding cylinder at midrapidity, we have
a disk with higher temperature in the interior surrounded by an annular
ring
in which the plasma is in the mixed phase at $T_c$, referred to as the $M$
region. Hadrons and hadronic clusters are formed in the $M$ region; the
hotter plasma in the interior feeds the $M$ region with quarks and keeps it
at $T_c$. The interior region shrinks as the temperature is lowered in
time, and the $M$ region shrinks as the hadronic clusters are emitted from
the boundary, even though quarks and hadronic clusters all move radially
outward.
\medskip
Restricting our attention to only the $M$ region, we map the 2D annular
ring
to a 2D square lattice of size $L\times L$ initially. We impose periodic
boundary condition in the vertical $y$ direction, and regard the first
column at $x=1$ as the inner boundary of the ring, and $x=x_y(t)$ as the
outer boundary (depending on $y$) that changes with time $t$, starting with
$x_y(0)=L$ for all $y$. The dependence of $x_y(t)$ on t will be a result
of the process of cluster emission from the plasma and a smoothing
process to be described below.
\medskip
Since we attempt to model the phase transition process as realistically as
possible, but without knowing all the dynamical mechanisms from first
principles, our rules will be elaborate by the standards of SOC, yet
rather simple from the physical point of view. Instead of writing down a
set
of differential equations in the continuum limit that completely specify
the
dynamics, which we know very little, we describe the cluster formation
process step-by-step in a series of rules. These rules are presented here
as a starting point, but can be modified as the description of the process
is varied or improved.
\medskip
\noindent (a) {\it Space-time.}\quad The spatial coordinates $(x,y)$ are
discretized into $L\times L$ units initially. We continue to refer to the
lattice space as the $M$ region, which will change in size with time. In
each time step all the  subprocesses to be described below are completed
before another step is taken. However, since the subprocesses involve
mainly
near-neighbor movements, no assumption of infinitely slow evolution rate is
assumed.
\medskip
\noindent (b) {\it Single-site nucleation.}\quad Initially, all sites are
unoccupied, representing quark phase. Let each site have the probability
$p$
of being occupied in each time step, representing a local hadron phase. The
initial hadron (not yet the type to be detected later, since it is at
$T_c$)
is thus of  size $S_0$, which is slightly less than the intersite distance,
so that neighboring sites can both be newly occupied without forming a
cluster.  In brief, we denote single-site nucleation (SSN) as $S_0=1$.  At
the next time step only the unoccupied sites can become occupied with
probability
$p$ at each site again.
\medskip
\noindent (c) {\it Growth.}\quad If a newly occupied site is a nearest
neighbor of an old occupied site, we regard it as a growth process, and
require the two to be bonded to form a cluster (or an extension of a
cluster, if the old site is already part of an existing cluster). All sites
in a cluster are bonded. Two neighboring sites need not be bonded except in
this growth process; in particular, two newly nucleated sites by themselves
are not bonded, as stated in (b) above.
\medskip
\noindent (d) {\it Average drift.}\quad All hadrons drift toward the outer
boundary at a constant velocity, simulating the radial expansion. This
drift motion in the $M$ region is represented by the requirement that all
occupied sites $s_i(t)=(x_i,y_i)$ are to move to $s_i(t+1)=(x_i+1,y_i)$ at
the next time step, assuming no fluctuations. A vacated sites becomes
unoccupied.
\medskip
\noindent (e) {\it Random walk.}\quad Regarding the clusters
as massive colloids in a fluid of the quark-gluon plasma, we require that
they take random walks around their average drift as in Brownian motion.
Thus if an occupied site in a cluster is at $(x_i, y_i)$, it is moved at
the
next time step to one of the following four possible
sites, $(x_i+2, y_i), (x_i+1, y_i+1),$
 $(x_i+1, y_i-1),$  and $(x_i, y_i)$, with equal probability. The average
is $(x_i+1, y_i)$ as stated in (d). All the other sites in the cluster move
together as a rigid body.
\medskip
\noindent (f) {\it Coalescence.}\quad As the clusters move, they may
collide. A collision occurs during the random walk when there is an overlap
of occupied sites belonging to two different clusters. When that happens,
we
require that the two clusters are treated as one thereafter, with one of
the
double at the overlap site be taken to occupy the nearest unoccupied site
(at
random if more than one), and be bonded to the rest of the enlarged
cluster.
\medskip
\noindent (g) {\it Boundaries.}\quad Since the quark phase on the inside of
the annular ring supplies the quarks to the mixed region, we require that
the quark density in the $M$ region remains constant throughout, including
the boundary at $x=1$. That means that all unoccupied sites remain
unoccupied until an occupied site moves there, or a nucleation takes place
there. Unoccupied sites do not drift.  At the outer boundary hadron
clusters
can leave the
$M$ region, but not the quarks due to confinement. Let a perimeter site be
defined to be the site to the immediate right of the site in the $M$ region
with the largest $x$ value for every fixed $y$, the latter being either
occupied or unoccupied.  When a cluster of $S$ sites breaks through the
outer
boundary, defined by any of the $S$ sites occupying a perimeter site, then
the whole cluster is removed from the
$M$ region, which is now redefined to have $S$ sites less, having a
transient boundary.
\medskip
\noindent (h) {\it Smoothing the outer boundary.}\quad By removing $S$
sites from the $M$ region, the transient boundary can become very irregular
due to the ``hole" created. Assuming positive surface tension, we smooth
the
transient boundary by filling the hole with nearby unoccupied sites until
a vertical boundary is achieved at the location of the hole. It cannot
extend to the entire height $L$, since it would depend on the value of $S$.

Obviously, if $S$ is small compared to $L$, there will be an indentation of
only $S$ sites one column to the left of the previous boundary line. If $S$
is not small, then filling the hole with $S$ unoccupied sites may expose
other occupied sites near the boundary, in which case the affected cluster
must also be emitted, resulting in further smoothing. This should be
continued until all existing clusters are within the new real boundary,
before another time step is taken.
\medskip
\noindent (i) {\it Evolution.}\quad After all clusters have taken their
walks, have enlarged due to possible coalescence, and have been removed
from the $M$ region upon crossing the outer boundary, which is then
smoothened, we initiate another round of nucleation at the
remaining unoccupied sites with the same probability $p$. If it occurs at a
site next to an existing cluster, it is to be bonded to that cluster as
described in (c). Then we let each cluster take a step in random walk and
repeat the process, again and again until all clusters are emitted from the
plasma. The evolution stops when there are no more unoccupied sites left,
and the phase transition is over.
\medskip
\noindent (j) {\it Cluster-size distribution.}\quad With $n_j(S)$ denoting
the number of clusters of size $S$ emitted in the $j$th event, and
$N_j=\sum_S\, n_j(S)$, we define the average distribution after
$\cal N_{\rm evt}$ events by
$$P(S) = {1\over {\cal N}_{\rm evt}}\
\sum_{j=1}^{\cal N_{\rm evt}}\ {n_j(S)\over N_j}.\eqno(1)$$
Our model calculation of cluster size $S$  may or may not coincide with the
experimental determination of $S$, since a cluster at the boundary of the
plasma is at $T_c$, while the detector measures individual hadrons in $T=0$
vacuum.  It is an issue that is outside the scope of the present
investigation. However, it is not unreasonable to expect that if there is a
scaling behavior in
$P(S)$, the scaling index may be identical in the two situations. In any
case, (1) should be used for the experimental definition of $P(S)$,
whatever
algorithm is used for the identification of a cluster.
\medskip
The above rules form the simplest set that can adequately describe the 2D
problem at hand. Note that there is no breakup, and that coalescence
occurs with 100\% probability when two clusters collide. Thus there are no
parameters to adjust that affect the nature of the outcome of a
collision. The only obvious parameter in the problem is $p$, which
summarizes
all the dynamics of nucleation. It is, however, not under experimental
control, so it cannot be used to tune the system to the critical point.
For every $p$ we shall carry out the simulation and calculate $P(S)$, and
then vary $p$ to check the dependence  of the result on $p$.  Ordinarily,
the lattice size $L$ is not a parameter in a lattice calculation in the
sense that one usually chooses $L$ to be as large as possible, given one's
computer capability.  But since our initial nucleation size $S_0$, which
has a physical meaning, is set equal to one unit of lattice spacing, $L$ is
therefore the initial size of the $M$ region in units of $S_0$, and cannot
be
arbitrary.  Results of the calculation for different values of $L$
correspond to different physical conditions in the heavy-ion collision.
\medskip
Based on  the above rules we have performed the calculation for $L=16$ for
various values of $p$. The results after $10^3$ events of simulation with
$p$ ranging from 0.01 to 0.5 are shown in Fig.\ 1.  Evidently, there is a
scaling behavior
$$P(S) \propto S^{-\gamma}\eqno(2)$$
for $p\geq 0.05$.  For smaller values of $p$ there is not enough time for
the clusters to grow to large sizes to exhibit a scaling behavior. This is
a finite-size effect for such low nucleation rates only. For higher $p$ not
only can the clusters grow faster, but also there are more clusters around
to enhance the collision frequency. The straightline behavior in the
log-log plot then becomes more manifest, and the slope $\gamma$ is clearly
identifiable.
\medskip
The finite-size effect can be checked by increasing $L$. In Fig.\ 2 the
results for $L=32$ are shown for the same set of $p$ values. The linear
regions are clearly extended for all $p$. Thus given enough time for growth
all cases exhibit scaling property. Since no parameters have been tuned
to bring the system to exhibit that behavior, this result may be regarded
as
evidence for SOC.
	\medskip
To see the dependence of $\gamma$ on $L$, we show in Fig.\ 3 a comparison
of the two cases $L=16$ and 32 for $p=0.2$. In the linear portion the two
distributions essentially overlap. The corresponding slope gives
$$\gamma=1.85.\eqno(3)$$
Thus there is independence on
$L$, which means independence on the initial size of the $M$ region. That
is
of physical signficance, since it implies that under different collision
conditions there is universality
 in the value of the scaling index $\gamma$. From Figs.\ 1 and 2 one can
see
that this independence on $L$ is valid also for other values of $p$.
\medskip
The dependence of $\gamma$ on $p$ is also not significant, although not
entirely negligible. For
$p$ ranging over a factor of 10, from 0.05 to 0.5, the value of $\gamma$
changes from 2.04 to 1.68, which is a 10\% deviation from the mean value,
i.e.,
$$\gamma=1.86\pm 0.18.\eqno(4)$$
Thus within that accuracy we may regard $\gamma$ as being independent of
$p$. In that context we may state that the scaling behavior obtained so far
is broadly universal, independent not only of the conditions under which
the quark-gluon plasma is formed, but also of the dynamics of nucleation
parametrized by $p$. That is, of course, of great phenomenological
significance, since it would allow experimental verification of the
property found here with few restrictions.
\medskip
It is of interest to ask what happens in the limit of large $p$. If $p=1$,
all sites would be occupied in the first step. One might then think that it
would lead to one large cluster, and $\gamma$ would become a large negative
number. However, that is not the case because neighboring occupied sites
are
not bonded until they overlap. While large clusters are in the process of
being formed by successive overlaps, some of the low $S$ clusters will
have left the $M$ region and populate the low $S$ portion of the $P(S)$
distribution.  In Fig.\ 4 we show the result for $p=0.9$ for $L$=16 and 32.
One first notices that there is no increase at large $S$.  Roughly, they
follow the same trend as for lower $p$, but there is no linear region and
the two distributions do not overlap.  For $L=16$ the curve is imbedded in
the range covered by the curves in Fig.\ 1, but for $L=32$ and $S>2,
 P(S)$ lies above all the curves in Fig.\ 2.  It suggests that for such a
large $p$ large clusters can form if given enough time, but sufficient
numbers of small clusters are always emitted in the early phase of the
transition to yield a distribution that is not too far off from the scaling
result of Figs.\ 1 and 2.  We shall not consider these unphysically large
values of $p$ again in the following.
\vskip.7cm
\noindent {\bf 3. DOUBLE-SITE NUCLEATION}
\medskip
There is one more aspect of the cluster formation problem that we have not
varied, {\it viz.}, the critical radius of nucleation. In the previous
section we considered single-site nucleation (SSN), which is equivalent to
setting
$S_0$ at essentially one unit of lattice spacing. Now let us consider a
larger nucleation radius without changing anything else. We do it by
requiring double-site nucleation (DSN), which turns out to necessitate some
change of rules.
\medskip
We replace rule (b) by

\noindent (b') {\it Double-site nucleation.}\quad Let each unoccupied site
still have probability $p$ of being occupied in each time step, but if it
happens, its occupancy is to be regarded as virtual. When two virtual
sites are nearest neighbors, then they become two bonded real
sites within the same time step.  It is a newly created hadron.  Three or
more virtual sites in an aggregate are bonded in pairs only as new hadrons
next to one another, but not as a big cluster. The odd one left out after
pairing (done randomly) is discarded.  A single virtual site that is a
near-neighbor to an existing cluster is bonded to that cluster in a growth
process, as in (c). However, if a single virtual site stands alone in an
environment of unoccupied sites, then nucleation fails, and we evict the
occupant and declare the site unoccupied.
\medskip
This rule simulates the failure of a hadronic bubble with radius less than
the critical radius to realize its hadronization. Our initial hadron is a
two-site ``cluster", whose shape should not be taken to have any physical
significance (e.g., a color-singlet bound-state); it is merely the simplest
nontrivial configuration. With this rule the simulation can proceed as
before. Note that nucleation now takes place with probability $p^2$,
while growth of an existing cluster at a neighboring site can take place
with probabililty $p$. This is more in accord with the physical reality
that growth is easier than nucleation.
\medskip
There is a small problem when the evolution is near the very end. If there
is
an isolated unoccupied site in the first column at the inner boundary, it
can
never become occupied, and the hadronization process will never be
completed. Thus we must add one more rule.
\medskip
\noindent (k) {\it Contracting the inner boundary.}\quad Whenever an
unoccupied site at $x=1$ along the inner boundary becomes isolated (without
occupied or unoccupied sites as neighbors), it must be moved to the nearest
site that is next to another unoccupied site, before another time step is
taken. If it is the last unoccupied site, it is discarded.
\medskip
This rule effectively eliminates some nonsites from interferring with the
nucleation process.  Instead of being arbitrary, it actually mimics reality
to a certain extent. The mapping of the annular ring (that becomes a disk
near the end) to a square lattice becomes singular when the inner quark
phase
shrinks to zero, so removing the nonsites from our consideration softens
that
singularity somewhat.
\medskip
The produced clusters can be measured in terms of their cluster sizes $S$,
as before. However, since the initial size $S_0$ is now doubled, we can
consider a normalized cluster size, $S'$, measured in units of $S_0$:
$$S'=S/2.\eqno(5)$$
It is the comparison of $P(S')$ in this case (DSN) with $P(S)$ in the
previous case (SSN) that is meaningful, since the growth of a cluster size
should be measured relative to its initial size. In the following we shall
always use the normalized size $S'$, but with the prime omitted, so only
$S$
will  appear in its place. It means that there will be
half-odd-integer values of $S$. The normalization of $P(S)$ remains to be
$$\sum_S  P(S) = 1,\eqno(6)$$
where the sum over $S$ now runs over integers and half-odd integers.
\medskip
In Figs.\ 5 and 6 are shown our results for DSN ($S_0=2$) for $L=16$ and
32,
respectively. Their general features are similar to those shown in Figs.\ 1
and 2, except for the oscillations at small $S$ and large $p$. The latter
can be understood as follows. At large $p$ there can be many DSN taking
place. Their collisions with one another lead to increases in cluster sizes
involving even number of sites. In the beginning growth by coalescence is
more efficient than growth by bonding single neighbors one at a time near
the outer boundary, the latter mechanism contributing to odd numbers of
total sites in a cluster. That is why there is a dip at $S=1.5$ for
$p=0.5$.
At large enough $S$ the clusters are emitted later in the evolution, so
both mechanisms will have enough chances to operate to smooth out of the
distribution.
\medskip
Let us ignore the oscillation and consider a line through the middle
representing the average distribution. We see then that the results in
Figs.\
5 and 6 again show scaling for all values of $p$ except
$p=0.01$, just as in the SSN case. The slopes of
the linear portion again do not depend on $p$ sensitively. Fig.\ 7 shows
the
comparison of the two cases $L=16$ and 32 for $p=0.2$. The dependence on
$L$
is evidently negligible in the scaling region.
\medskip
To provide a better comparison of the results in the two cases, SSN and
DSN, we select the $p=0.2$ distributions and exhibit them in Figs.\ 8 and 9
for $L=16$ and 32, respectively.  The reason why the dashed ($S_0=2$)
curves
are lower than the solid ($S_0=1$) curves is due to the normalization
condition (6). For the solid curves the sum is over integer values of $S$
only, while for the dashed curves the sum must include the half-odd
integers
as well. The difference has no effect on the slopes. Thus again $\gamma$ is
the appropriate characterization of the problem, and is now shown to be
independent of the nucleation radius $S_0$ also.
\vskip.7cm
\noindent {\bf 4. CONCLUSION}
\medskip
We have investigated the problem of cluster growth in heavy-ion collisions
when there is a quark-hadron phase transition, and found interesting
scaling behavior in the distribution of cluster sizes that exhibit
universal properties. The two-dimensional approximation of the geometry of
the problem is not an unreasonable oversimplification, if one restricts
one's observational domain to a small rapidity interval at midrapidity.
The square lattice approximation may distort the result somewhat, but the
general features of the result should remain unchanged.
\medskip
We have found that for $p$ in the range from 0.05 to 0.5 the scaling index
$\gamma$ is around 1.9.  It is independent of the size of the $M$ region,
which in turn is dependent on the  collisional conditions of the heavy
ions, assuming that a quark-gluon plasma is formed.  Since from collision
to
collision the impact parameter changes and the initial temperature may
vary,
it is crucial that we focus on some quantity that is insensitive to those
experimental conditions.  It is also important that the quantity is
insensitive to those parameters that are unknown or poorly known in theory,
such as the critical nucleation radius and nucleation rate. The
independence
of $\gamma$ on $S_0$ and its approximate independence on $p$ further
promote
$\gamma$ as an observable that can possibly survive the experimental and
theoretical
uncertainties to serve as a numerical characterization of the phase
transition process.
\medskip
We have referred to the problem as one in quark-hadron phase transition,
which carries the usual connotation that the system under study is in
thermal equilibrium.  Indeed, we have set up the problem in the familiar
framework built on the hydrodynamical description of an expanding
cylindrical fluid with a temperature profile.  We then focused our
attention on the mixed phase, which again is defined in the conventional
framework of equilibrium thermodynamcis.   However, the heart of the
cellular automata does not rely crucially on the validity of those
concepts.  If thermal equilibrium were not achieved in heavy-ion collision,
we would still have the problem of hadronization of a dense quark-gluon
medium, and the rules that we have adopted to simulate what may happen may
still be applicable with little modification.  Thus by working with
cellular automata without reference to such quantities as free energy,
temperature, pressure, equation of state, etc., the validity of our
investigation may transcend that of  the
study of quark-gluon plasma based on the usual hypotheses.
If the physics involved can be further illuminated by recognizing it as a
process in self-organized criticality, perhaps deeper understanding will be
forthcoming from an unexpected quarter in theoretical physics.
\medskip
We, of
course, do not know from first principles whether our description of the
cluster growth process is correct in nature.  What is presented here is a
first step in the direction of asking questions that cannot at the moment
be
easily posed in the continuum approximation of the complicated system, let
alone answered.  Ultimately, it is the experimental verification of what we
have found here that can give credence to our description of the dynamical
process.  Such experimental investigations are not likely to be
contemplated, if there is no focus on what to look for or hint of what
might
be in store.  In that respect it is hoped that our finding of a universal
scaling index may provide the needed impetus.
\vskip.7cm
\centerline{\bf Acknowledgment}
\medskip
We have benefitted from helpful comments by T. Hwa on our earlier work.
This work was supported in part by the U.S. Department of Energy under
Grant No. DE-FG06-91ER40637, and by the Natural Science and Engineering
Council of Canada and by the FCAR fund of the Quebec Government.
\vfill

\eject
\centerline{\bf REFERENCES}
\bigskip
\item{[1]\quad} R.C. Hwa, C.S. Lam, and J. Pan, Phys. Rev. Lett. {\bf 72},
820
(1994).
\medskip
\item{[2]\quad} P. Bak, C. Tang, and K. Wiesenfeld, Phys. Rev. A {\bf 38},
364
(1988).
\medskip
\item{[3]\quad} J.M. Carlson, J.T. Chayes, E.R. Grannan, and G.H. Swindle,
Phys. Rev. Lett. {\bf 65}, 2547 (1990).
\medskip
\item{[4]\quad} P. Bak, K. Chen, and C. Tang, Phys. Lett. A {\bf 147}, 297
(1990).
\medskip
\item{[5]\quad} P. Grassberger and H. Kantz, J. Stat. Phys. {\bf 63}, 685
(1991).

\vskip2cm
\centerline{\bf Figure Captions}
\bigskip
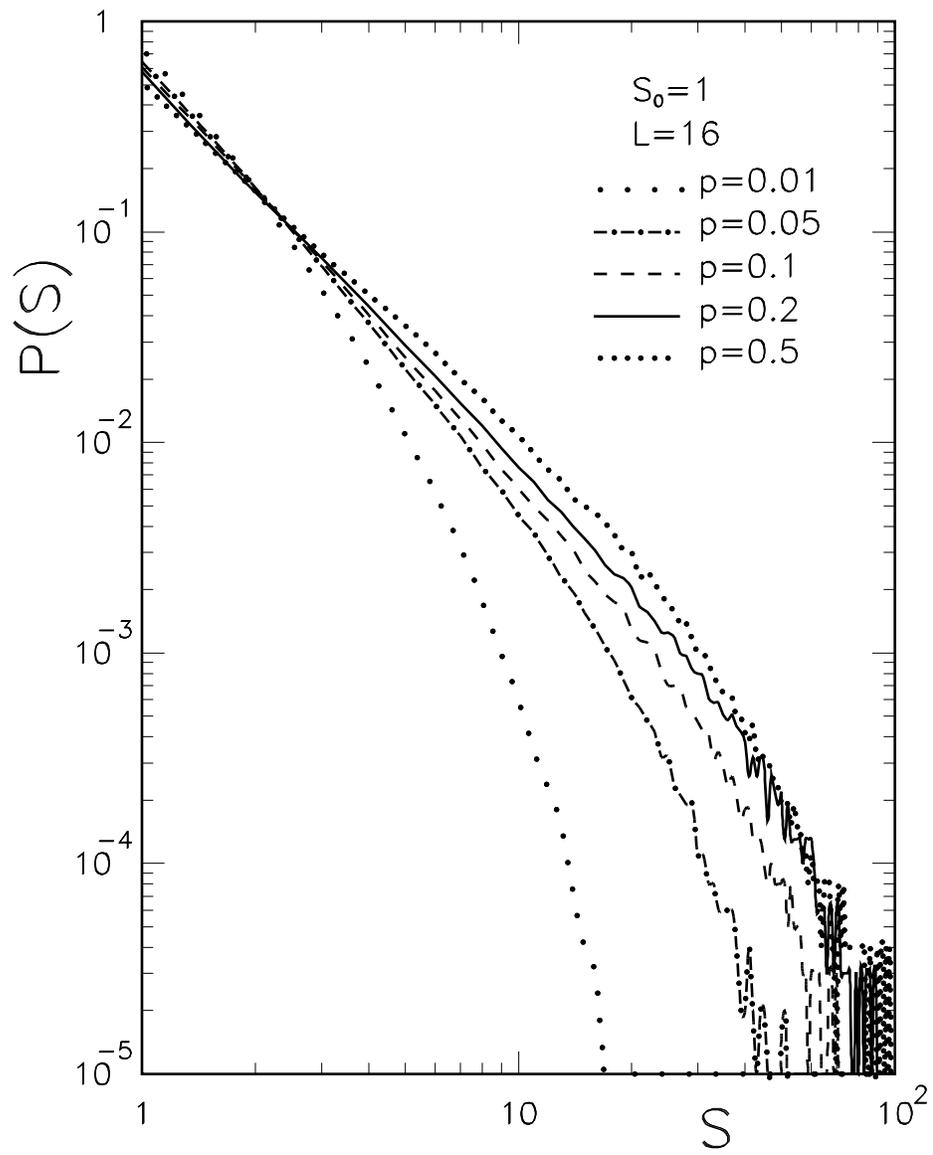
\item{Fig.\ 1.\quad} Cluster-size distribution $P(S)$ for single-site
nucleation (SSN) and $L=16$, and for various values of nucleation
probability
$p$.
\medskip
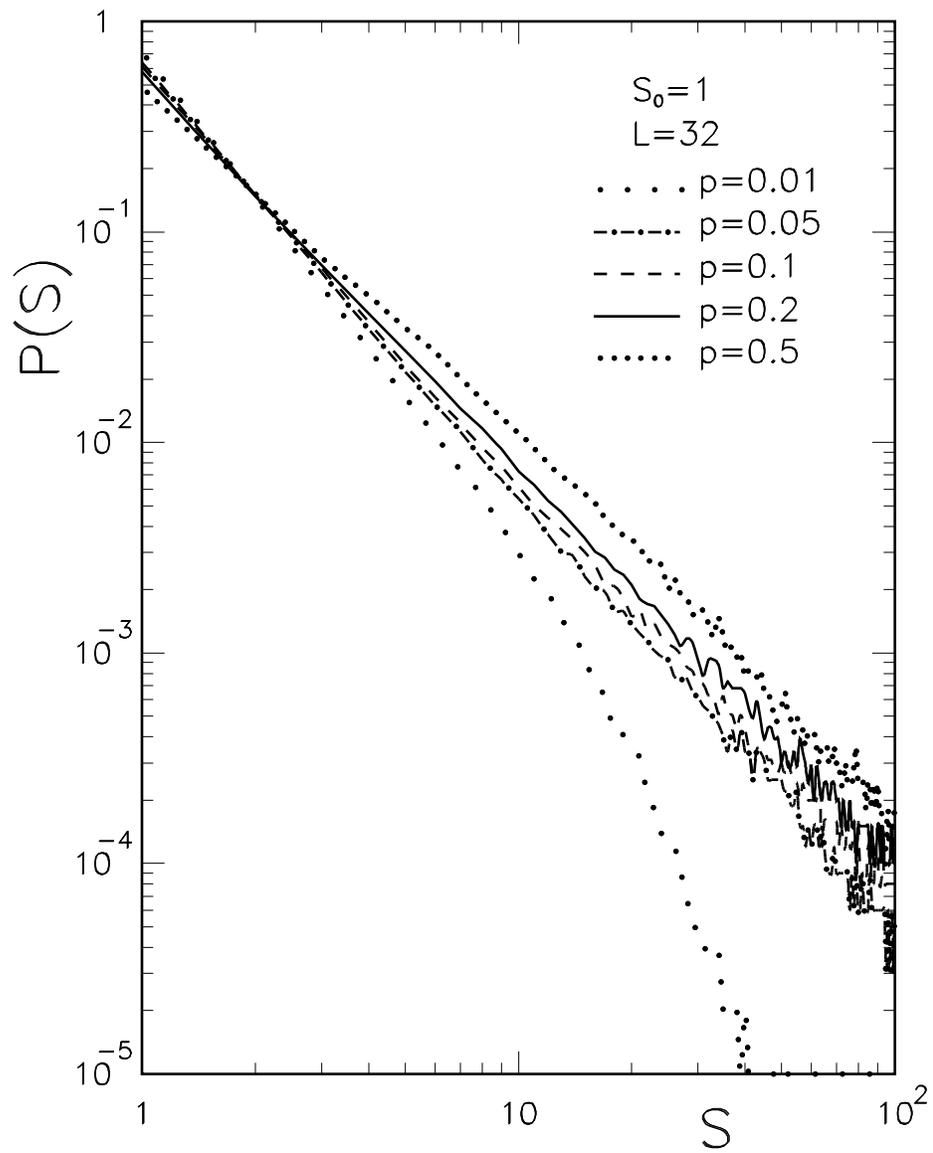
\item{Fig.\ 2.\quad} Same as in Fig.\ 1 but for $L=32$.
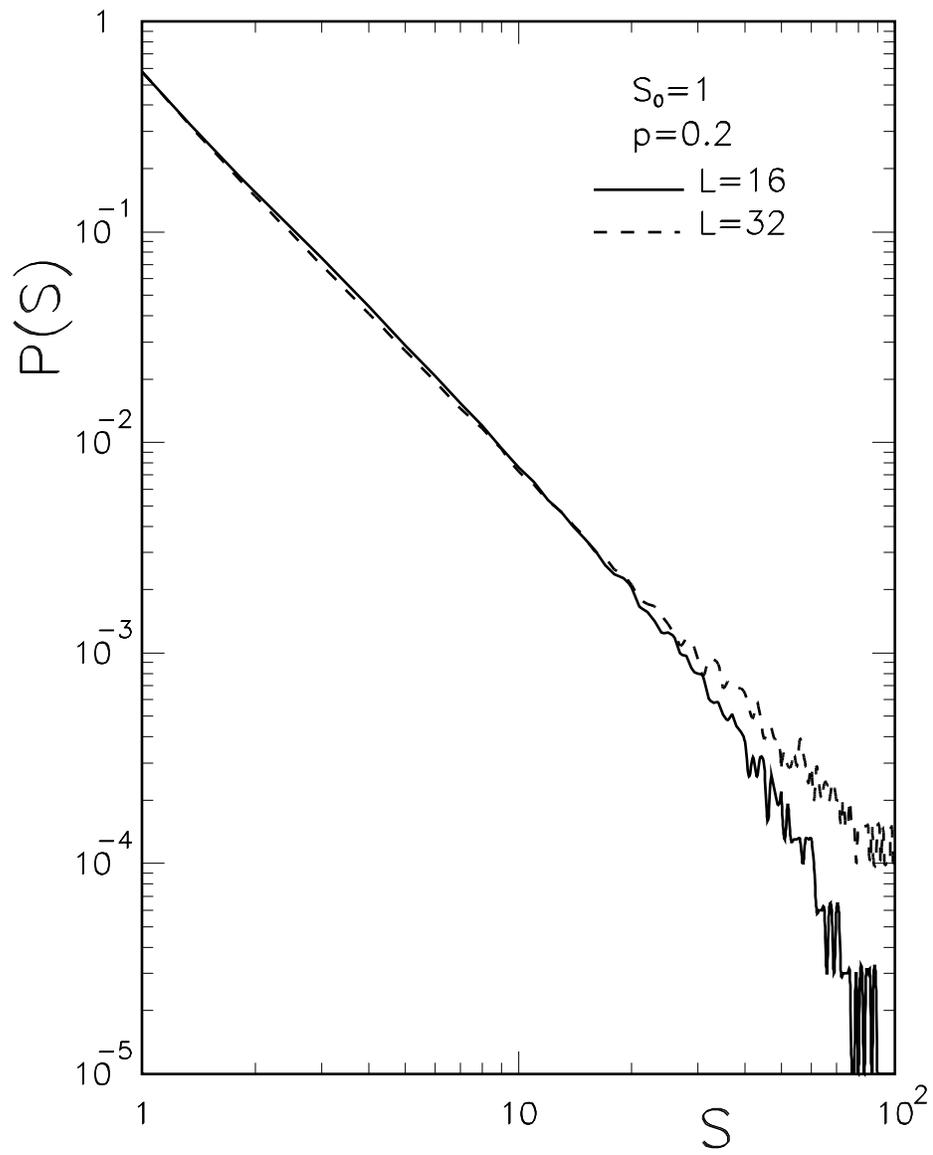
\medskip\item{Fig.\ 3.\quad} $p=0.2$ from Figs.\ 1 and 2.
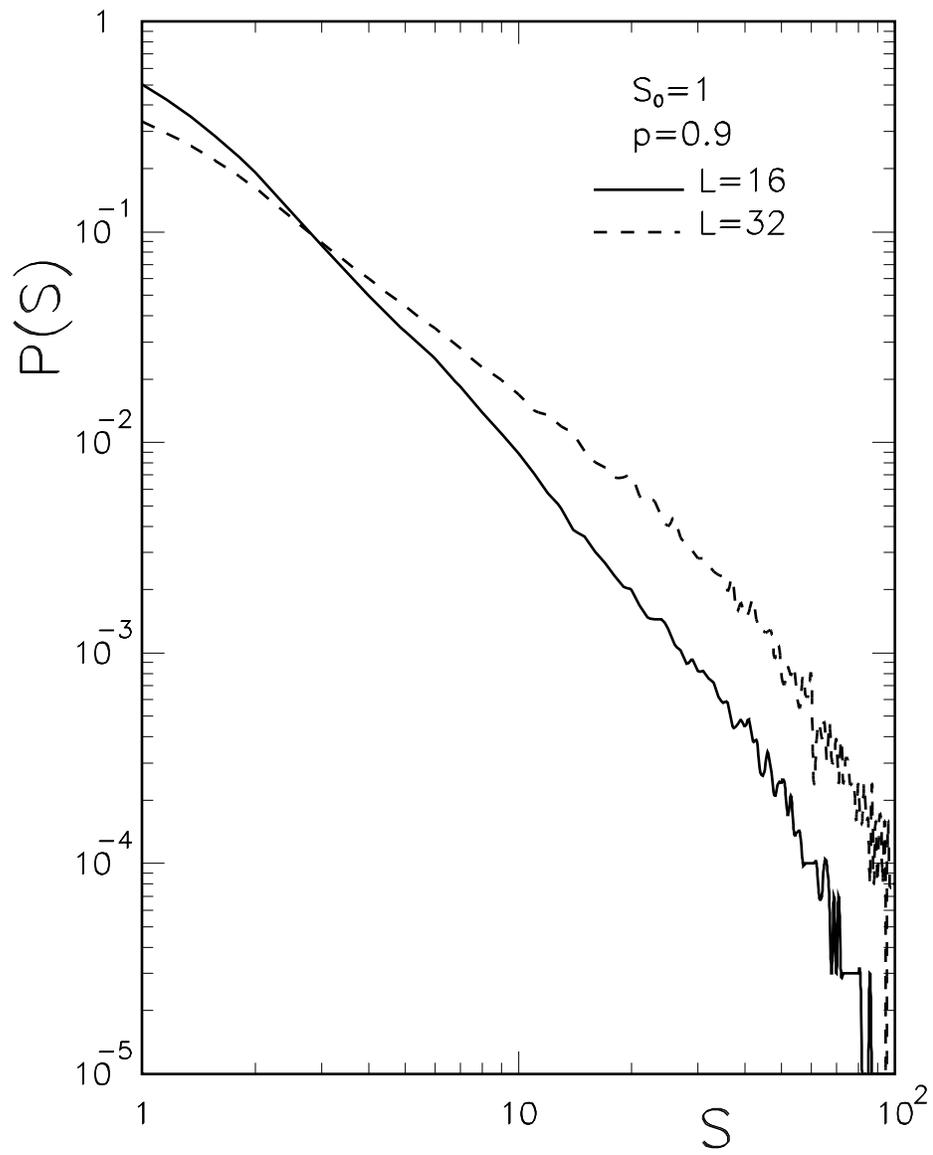
\medskip\item{Fig.\ 4.\quad} $P(S)$ for $p=0.9$.
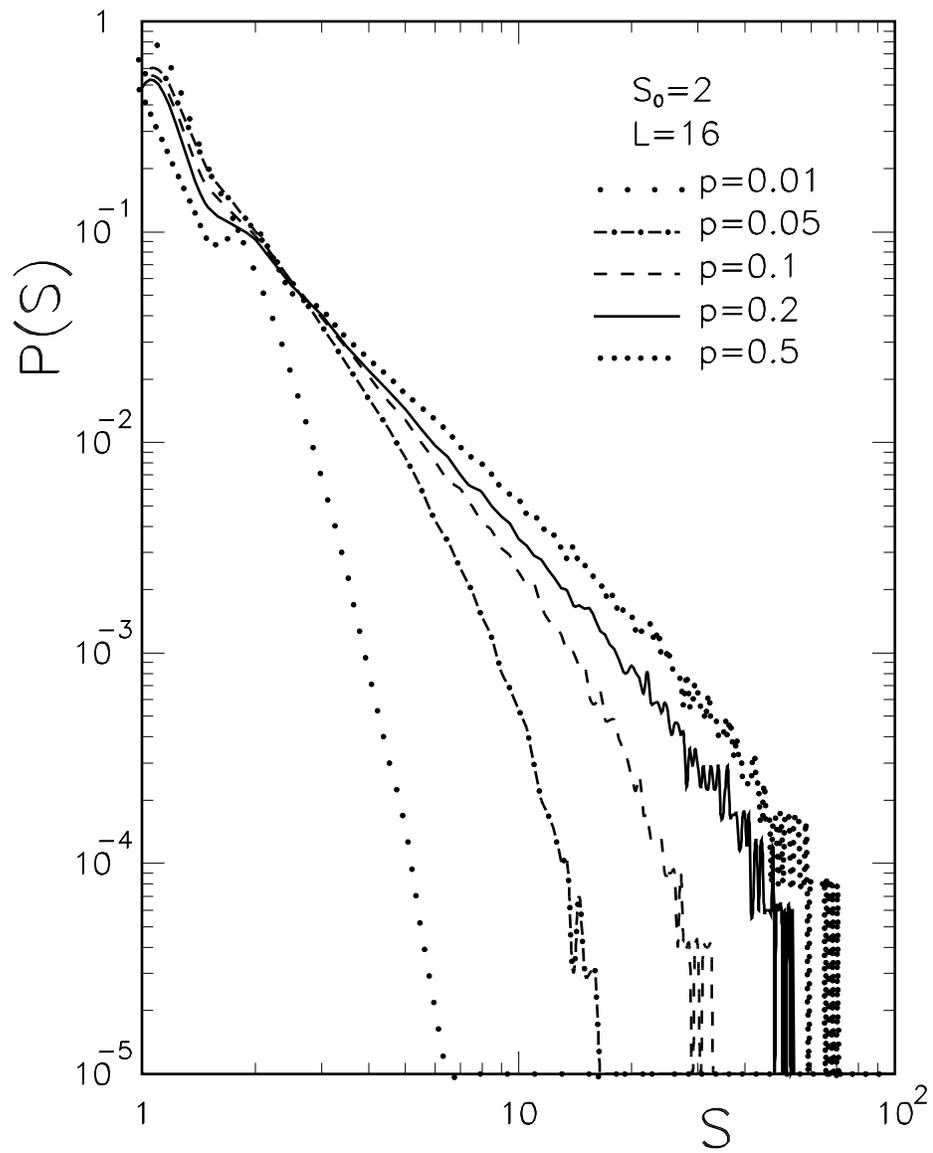
\medskip\item{Fig.\ 5.\quad} Same as in Fig.\ 1 but for double-site
nucleation (DSN).
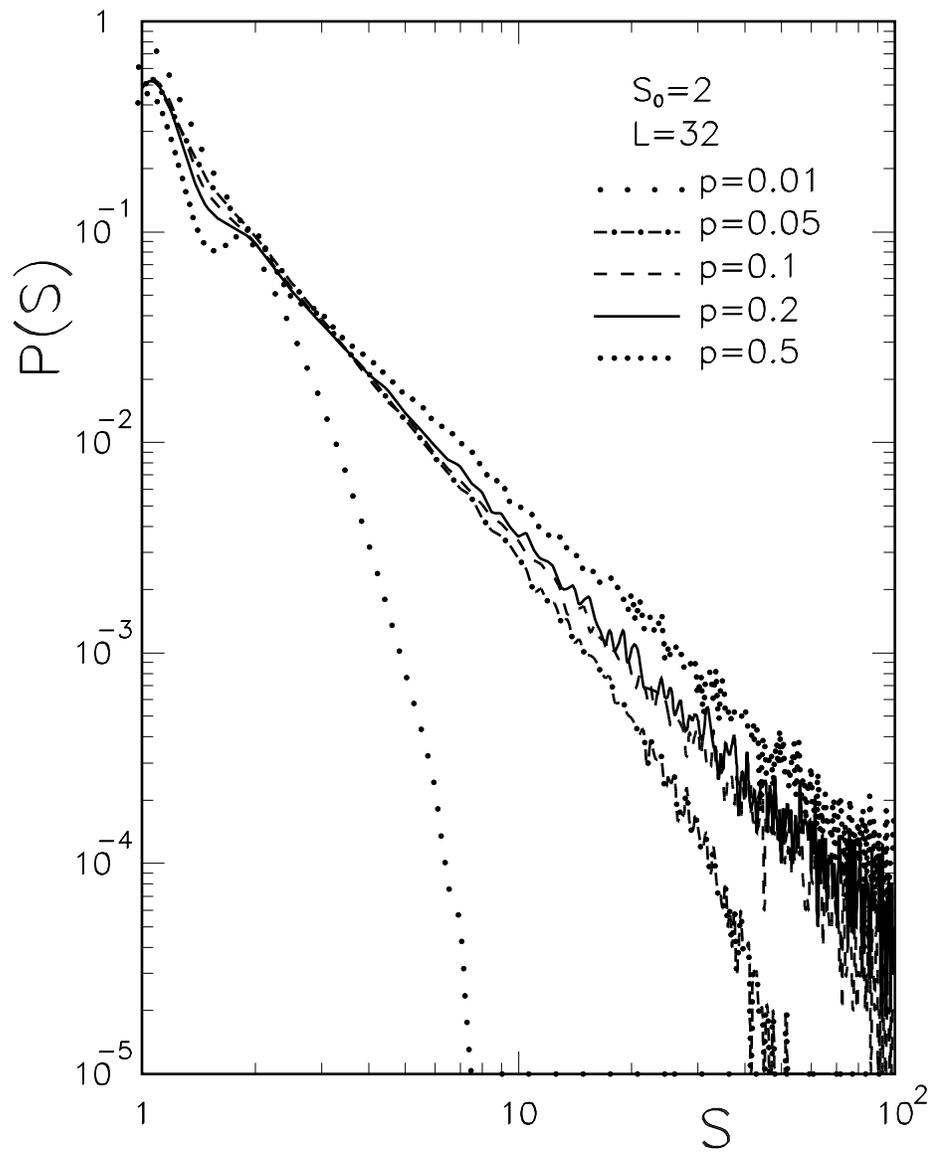
\medskip\item{Fig.\ 6.\quad} Same as in Fig.\ 2 but for DSN.
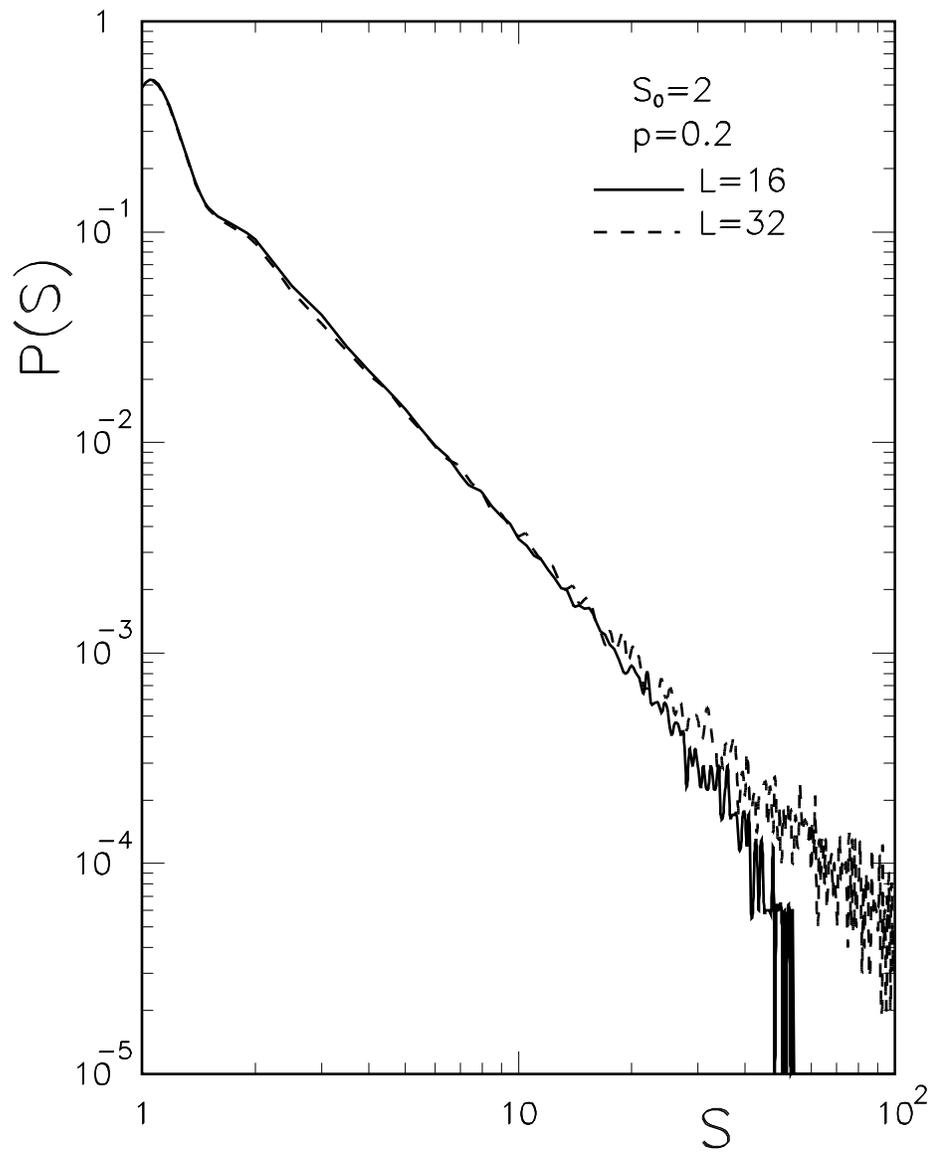
\medskip\item{Fig.\ 7.\quad} $p=0.2$ from Figs.\ 5 and 6.
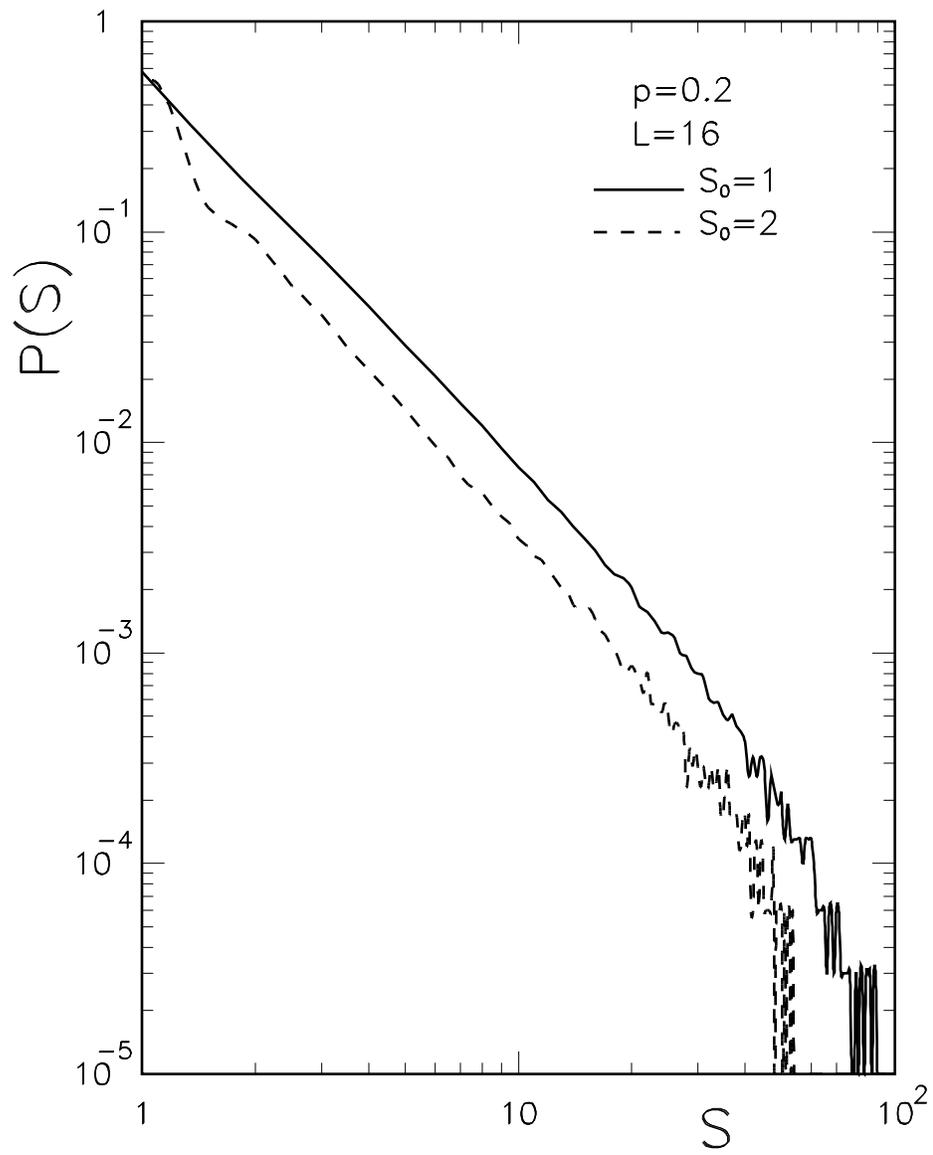
\medskip\item{Fig.\ 8.\quad} $p=0.2$ from Figs.\ 1 and 5.
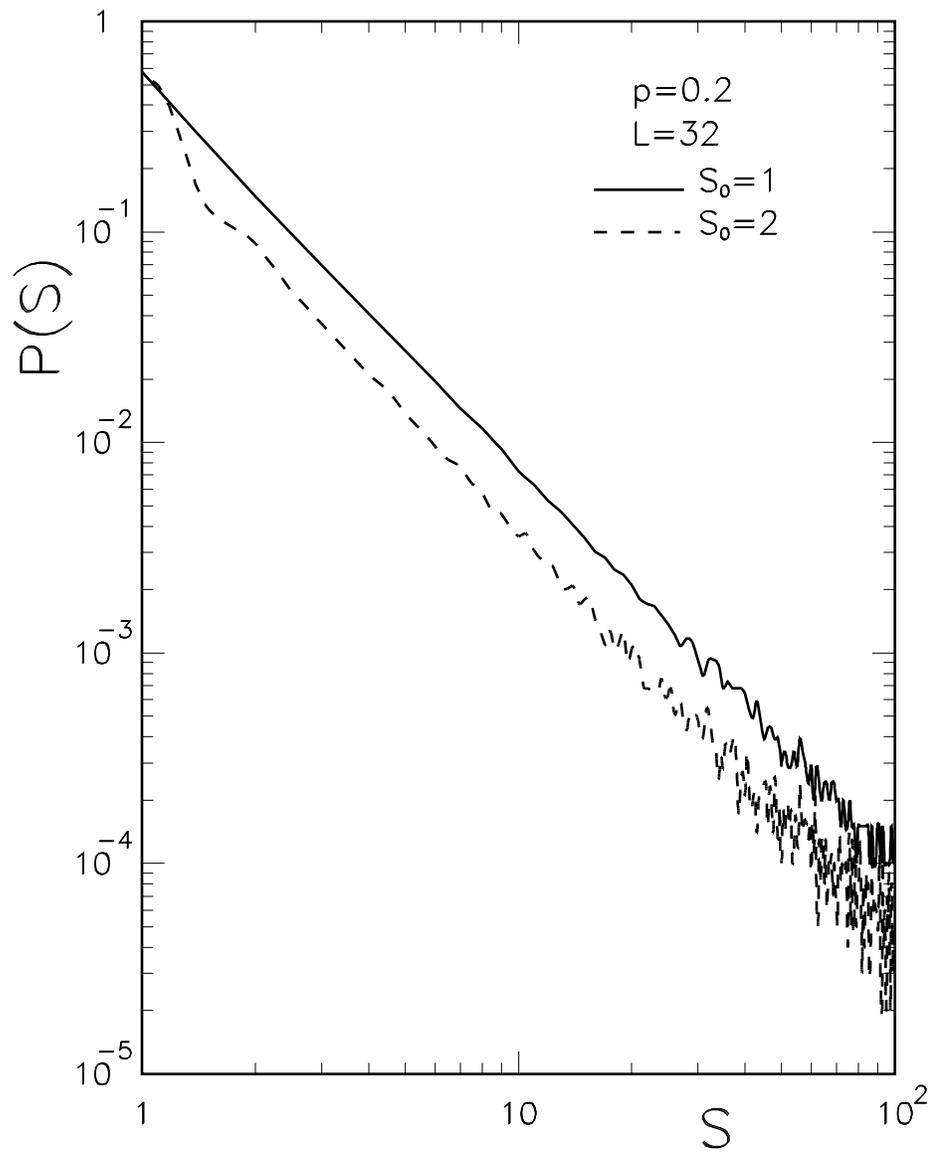
\medskip\item{Fig.\ 9.\quad} $p=0.2$ from Figs.\ 2 and 6.

\end